\documentclass[preprint,showpacs,preprintnumbers,amsmath,amssymb]{revtex4}

\usepackage{graphicx}
\usepackage{dcolumn}
\usepackage{bm}

\begin{document}
\topmargin=0.1cm

\preprint{MIT-CTP-3374, LAUR-03-3581}

\title{Dressing up the kink}

\author{Yoav Bergner}
 \email{bergner@mit.edu}
\author{Lu\'{\i}s M. A. Bettencourt}
 \email{lmbett@lanl.gov}
  \homepage{http://www.mit.edu/~lmbett}
\affiliation{
Center for Theoretical Physics, 
Massachusetts Institute of Technology,
77 Massachusetts Avenue, Cambridge MA 02139 USA }
\affiliation{Los Alamos National Laboratory, MS B256,
Los Alamos NM 87545}

\date{\today}

\begin{abstract}
Many quantum field theoretical models possess non-trivial solutions
which are stable for topological reasons.  
We construct a self-consistent example for a self-interacting scalar
field--the quantum (or dressed) kink--using a two particle irreducible effective action in the Hartree approximation. This new solution includes quantum fluctuations 
determined self-consistently and nonperturbatively at the 1-loop resummed level and allowed to
backreact on the classical mean-field profile. This dressed kink is static under the 
familiar Hartree equations for the time evolution of quantum fields.  
Because the quantum fluctuation spectrum is lower lying in the presence of the 
defect, the quantum kink has a lower rest energy than its classical
counterpart.  However its energy is higher than well-known strict
1-loop results, where backreaction and fluctuation self-interactions are omitted. 
We also show that the quantum kink exists at finite temperature and that its profile broadens 
as temperature is increased until it eventually disappears. 
\end{abstract}

\pacs{03.70.+k, 05.70.Ln, 11.10.-z, 11.15.Kc \hfill}

\maketitle

Topological defects are perhaps the best known nonlinear solutions of
quantum field theories.  They are explicitly spatially inhomogeneous yet localized and have
finite energy configurations (sometimes involving
defect-antidefect pairs) \cite{Rajaraman,Vilenkin}. 
Moreover topological defects, particularly extended ones (strings, domain walls), 
have configurational entropies that  increase exponentially with their size. 
For these reasons defects are important vehicles of disorder. They can lead to phase 
transitions not predictable in their absence, such as Kosterlitz-Thouless \cite{Kadanoff}, 
or become the mesoscopic traces of high energy disorder, as happens in 
defect formation under sudden quenches. Defects can also lead to nonperturbative 
screening effects, e.g. of static magnetic fields by 't-Hooft-Polyakov monopoles \cite{Polyakov}. 

Because defects are localized solutions carrying quantized topological numbers 
they can often be thought of as generalized quantum particles. In fact in certain circumstances 
topological solutions dominate the phenomenology and should then be thought as the fundamental 
excitations of a new (often more tractable) model, related to the original by a duality transformation. 
Such changes of perspective have been the source of valuable insights into the nonperturbative 
properties of models of condensed matter, high-energy particle physics and quantum gravity.

While the existence and stability of topological solutions follows from general considerations 
about the symmetries and dimensionality \cite{Kibble} of a given model, the detailed properties 
of topological solutions, e.g. their size and energy, depend on the back-reaction from the 
background of quantum (and thermal) fluctuations in which they live.
For example the difference in energy between the dressed defect and its classical 
counterpart will affect estimates for the probability of its spontaneous nucleation 
via quantum or thermal fluctuations \cite{LytheHabib}. The same rest energy shift 
also affects the defect's dynamical response. Only a self-consistent treatment, where 
mass corrections occur to all orders in the loop expansion $\hbar \lambda$ can begin 
to probe the motion at non-zero momentum and long times \cite{BB1}.   
In some cases fluctuations may even
stabilize nontopological classical field profiles \cite{Jaffe}, but it
is not known whether 1-loop results survive a self-consistent treatment. 

Incorporating the effects of fluctuations is technically quite
difficult, which explains in part why most previous work has relied on
a strict loop expansion, neglecting either backreaction on the classical
mean-field profile or self-coupling of the fluctuation degrees of
freedom.  For example, calculations carried out to 2-loop order appear
in Ref.~\cite{2loop}, and a variational self-consistent calculation {\it without}
backreaction on the mean field is performed in Ref.~\cite{Boyanovsky}.
Since the corresponding ``solutions'' are not self-consistent, in
particular they will not be static under non-equilibrium quantum field
evolutions \cite{non-eq}.  Fermionic backreaction has been considered in
the calculation of the sphaleron energy, which is relevant for
estimating baryon number violation \cite{Moore,Schroers}.  This
computation showed the sphaleron energy to be quite insensitive to
such corrections,  however quartic self-couplings of the scalar field
were neglected \cite{Schroers}. It is precisely these nonlinear field
selfinteractions which we wish to consider here.

To start addressing the above issues we construct here an
example of a self-consistent quantum topological defect. For
simplicity, and for the benefit of analytical cross-checks on our numerical
results at both the classical and 1-loop quantum levels, we analyze
the familiar $\phi^4$ kink in 1+1 dimensions.  There has also been a
lattice Monte Carlo study of the quantum kink \cite{Ardekani}, providing 
a benchmark against which we can measure the efficacy of the 
self-consistent Hartree approximation. 
The methods developed here promptly generalize to other cases. 
Self-consistent spherically symmetric 
droplets, in particular, will be discussed elsewhere \cite{scbounce}.  
We include quantum effects at the 1-loop resummed level with
backreaction on the mean field by applying the familiar Hartree 
approximation to a $2PI$ effective action \cite{2PI}.  This technique places the
two-point correlation function on the same level as the one-point
function, or mean field, thus backreaction is included as naturally as
the nonperturbative resummation of fluctuations.  At this level of 
approximation, we obtain a local, nonlinear eigenvalue-type equation
for the quantum fluctuations. The next level of improvement, at 2-loop
resummed, would lead to a nonlocal integro-differential problem, which
would require methods of solution different from the ones developed below.  
Specifically we take the $\lambda \phi^4$ model with  potential
\begin{eqnarray}
V(\phi) = \frac{\lambda}{4} \left( \phi^2 - \frac{\mu^2}{ \lambda}  \right)^2.
\end{eqnarray}
Then the {\it classical} kink solution has the well known analytical form 
\begin{eqnarray} 
\phi(x) = \frac{\mu}{\sqrt{\lambda}} \tanh \left( \frac{\mu}{\sqrt{2}} x \right).
\end{eqnarray}

To set up the quantum problem we separate, as usual, the quantum field into a mean 
$\varphi \equiv \langle \phi  \rangle$ and a fluctuation operator-valued field $\hat \psi$ 
($\langle \hat \psi \rangle =0$), such that  $\phi(x,t)= \varphi(x,t) + \hat \psi(x,t)$.
The Hartree (or 1-loop resummed) equations for the mean field and the connected (Wightman) two-point function 
$G(x,t;x',t') = \langle \hat \psi(x,t) \psi(x',t') \rangle$ then become
\begin{eqnarray}
&& \left[ \Box - \mu^2 + \lambda \varphi^2(x,t) + 3 \lambda G(x,t) \right]
\varphi(x,t) = 0, \label{eqphi}  \\
 && \left[ \Box + \chi(x,t) \right] G(x,t;x',t') =0,  \label{eqG} \\ 
 && \chi(x,t) = -\mu^2 + 3 \lambda \varphi^2(x,t) + 3 \lambda G (x,t),
\label{chidef}
\end{eqnarray}
with $\Box = \partial_t^2 - \partial_x^2$ and $ G (x,t) \equiv
G (x=x',t=t')$.   We seek  a static solution to  equations (\ref{eqphi}-\ref{chidef}),
 obeying boundary conditions $\varphi(x\rightarrow + \infty) =
 -\varphi(x\rightarrow - \infty) = \varphi_0$, where $\varphi_0$ is
 one of the degenerate vacuum expectation values of the quantum
 field. These boundary conditions imply that the mean field must cross
 zero at some $x_0$, which defines the center of the kink.  

Eqs.~(\ref{eqG}-\ref{chidef}) can be converted into a (non-linear) eigenvalue problem by the familiar procedure of decomposing $G (x,t;x',t')$ in a complete orthonormal basis of mode fields, 
which we shall denote by $\{ \psi_k(x,t) \}$.  The procedure has been detailed extensively elsewhere 
\cite{non-eq}, and we do not repeat it here. The result is 
\begin{eqnarray}
G (x,t;x',t') = \sum_k \left\{ (N(k) + 1) \psi_k(x,t) \psi_k^*(x',t') + N(k) \psi_k^*(x,t) \psi_k(x',t') \right\}, 
\end{eqnarray}
where $N(k)$ is occupation number in this basis, which becomes  the 
Bose-Einstein distribution $n_B(k)$, if the system is in  thermal equilibrium 
 $N(k) \rightarrow n_B(k) = \left( e^{\hbar \omega_k/T} -1\right)^{-1}$, with $\omega_k$ the energy eigenvalue associated with the eigenvector $\psi_k(x,t)$. 
 In particular, the equal-point function  acquires the very simple form
\begin{eqnarray}
G (x,t) = \sum_k \left[ 2 N(k) + 1 \right] \psi_k(x,t) \psi_k^*(x,t).
\label{Geq}
\end{eqnarray}

Next we specify a basis $\left\{ \psi_k(x,t) \right\}$, together 
with the scalar product 
\begin{eqnarray}	
&& \left( \psi_k(x,t) . \psi_q(x,t) \right)= i \int d x \left\{ 
\psi_k^*(x,t) \partial_t \psi_q(x,t)  - \psi_q(x,t) \partial_t \psi_k^*(x,t) \right\}. \label{sproduct}
\end{eqnarray}
Because we seek a static solution the self-consistent  $\chi(x)$ is time independent. 
Then each $\psi_k(x,t)$ can be separated into the product of functions of time and space as  
\begin{eqnarray}
\psi_k(x,t) = \sqrt{ \frac{\hbar}{2 \omega_k}} e^{- i \omega_k t}  g_k(x),
\label{formpsi}
\end{eqnarray}
and the inner product (\ref{sproduct}) can be written in terms of the $g_k(x)$ as
\begin{eqnarray}
\left( g_k(x) . g_q(x) \right)  = \int dx~  g_k^*(x) g_q(x),
\end{eqnarray} 
which is familiar from quantum mechanics. With these definitions orthogonality in 
$\{g_k\}$ is equivalent to orthogonality in $\{\psi_k\}$. Moreover the completeness relation
\begin{eqnarray}
\sum_k g_k(x) g_k^*(x') = \delta^D(x-x'),
\label{completeness}
\end{eqnarray}
must be satisfied as a result of the canonical equal-time commutation relations obeyed 
by the original quantum field. Finally the explicit self-consistent eigenvalue problem reads 
\begin{eqnarray}
\left[  - \nabla_x^2 + \chi(x)  \right] g_k(x) = \omega_k^2 g_k(x) 
\label{sfeigen},
\end{eqnarray}
with the eigenmodes obeying periodic boundary conditions in the volume 
$x \in \left\{-L,L\right\}$:   $g_k(L) = g_k(-L)$, and 
solved in conjunction with equation (\ref{eqphi}) for $\varphi$.
In practice we discretize the fields on a spatial lattice of size $N$ and volume $L$.
This provides a regularization of the quantum field theory. In addition we have to  
devise a renormalization scheme to render all results finite as the lattice spacing 
$a = L/N$ is taken to zero, i.e. in the continuum limit.

The 2-point function $G $, as defined in   Eq.~(\ref{Geq}), contains an ultraviolet
logarithmic divergence in the continuum limit.  We specify the fluctuation physical mass 
and field expectation value over the trivial vacuum. These choices implicitly define the 
bare mass parameter $\mu^2$ as a function of the ultraviolet cutoff $\Lambda=\pi/a$  
so as to absorb this divergence. Explicitly 
\begin{eqnarray}
\label{renorm}
&& \chi_0= 2 m^2, \ \varphi^2_0=m^2/\lambda;   \quad   \rightarrow
 \quad  \mu^2 \equiv m^2  + 3 \lambda G_0, \\
&& {\rm where} \quad  G_0 = \int_{-\Lambda}^{\Lambda} \frac{dk} {2 \pi}  \frac{ \hbar} {2 \sqrt{k^2 +2 m^2}}.
\end{eqnarray}
Substitution of $\mu^2$, hence defined, into expression (\ref{chidef}) for 
$\chi(x)$  now leads to finite results, independent of $\Lambda$, as the continuum 
(and infinite volume limit) are taken.  

The only remaining divergences occur in the calculation of the energy.  The total energy can 
be split into a classical piece $E_{\rm cl}$ and the quantum (and thermal) fluctuation 
contribution $E_{\rm q}$. Taking expectation values we can write them as  
\begin{eqnarray}
E_{\rm q} && =  \int dx~ \sum_k \left\{ \psi_k(x) \left[ -\nabla^2 + \chi(x) \right] \psi(k) \right\} 
\left[1+ 2 N(k) \right] - \frac{3} {4} \lambda G^2(x)  \nonumber \\
&& \equiv \delta m  - \frac{3}{4} \lambda   \int dx  ~G^2(x)  + \sum_n \omega_n N(k_n),  \label{Eq} \\
E_{\rm cl} && = \int dx~ \left[  \frac{ (\nabla \varphi(x))^2}{2} +\frac {\lambda}{4} 
(\varphi^2 - \mu^2 /\lambda)^2  \right] . \label{Ecl}
\end{eqnarray}
The analysis of divergences is familiar at 1-loop, where it has been resolved in at least two
distinct ways:  by direct analysis of the divergences in a large
momentum expansion \cite{Dashen,Rajaraman} and via the normal ordering
of the fluctuation Hamiltonian operator, which is sufficient in 1+1 dimensions to extract the
physical finite contribution \cite{Cahil,BT,Weidig}.  Because they are implemented 
very differently we  have employed both methods as a check.

The most straightforward procedure, in our opinion, follows from direct analysis 
of the divergences in a large momentum expansion. For consistency we should 
start by substituting $\mu^2$ defined in (\ref{renorm}) into (\ref{Eq}-\ref{Ecl}), which   
eliminates all logarithmic divergences, up to an infinite constant. 
This constant results from the trace $\delta m = \sum_n \omega_n/2$. For large enough 
$n$, $\omega_n \rightarrow \omega^0_n = \sqrt{k_n^2 +2 m^2} \sim \vert k_n \vert$,
resulting in a leading quadratic divergence. This divergence is also 
characteristic of the trivial vacuum energy and should be cancelled against it.
With these prescriptions we can write the renormalized $E_q$ as   
\begin{eqnarray}
E_q^R =  \frac{1 }{ 2} \sum_n\left[ \omega_n \left( 1+2 N(k) \right) - \omega^0_n \right] 
- \frac{1}{2} G_0 \int dx \left[ \chi(x) - \chi_0 \right] 
- \frac{3}{4} \lambda   \int dx \left[G(x)-G_0\right]^2, \label{E2} 
\end{eqnarray}
The classical energy piece $E_{\rm cl}$ is now to be evaluated as in (\ref{Ecl}), with the substitution 
$\mu^2 \rightarrow m^2$.

Alternatively a finite energy expression for the trace in $\delta m$ can be 
constructed by suitably manipulating the fluctuation Hamiltonian operator. 
The normal ordered Hamiltonian is written in terms of a complete set of 
creation and annihilation operators in the kink background,  which are then 
transformed via a Bogoliubov transformation to the trivial vacuum basis, where 
vacuum expectation values are computed.  The residual contribution is finite 
and results from the existence of a non-zero particle number as 
seen back in the kink background mode basis. The final result is  
\begin{eqnarray}
\delta m = - \frac{1}{4} {\rm Tr} \left[ \frac{ ({\cal O} - {\cal O}_0)^2}{ {\cal O}_0 } \right],
\end{eqnarray}
where ${\cal O}= \left[- \partial_x^2 + \chi(x) \right]$ and  
${\cal O}_0= {\cal O}$ with $\chi(x) = \chi_0$, corresponding to the trivial vacuum. 
The advantage of this procedure is that it is independent of the exact form of $\chi(x)$ 
in ${\cal O}_k$ and is thus equally valid at strict 1-loop order as well as in the resummed 
self-consistent approximation. In practice we cast this expression in the explicit form 
\begin{eqnarray}
\delta m = - \frac{1}{4} \sum_{i,j} \vert  \left ( g_i (x) . g^0_j (x) \right) \vert^2 
\left[ \frac{ \omega^2_i }{\omega^0_j } - 2 \omega_i + \omega^0_j \right]. 
\label{E1}
\end{eqnarray}

There remains a final ambiguity in expressions (\ref{E2} - \ref{E1}), 
namely the prescription of the upper (ultraviolet) cutoff in the mode sums.  
In the presence of a non-trivial background one expects a spectrum that contains a 
finite and discrete set of bound modes, lying below a continuum of scattering states. 
Physically it is the advent of these bound states that produces the leading (in magnitude)
contribution to the energy shift of the quantum kink relative to the classical solution. 
Even for the lattice regularized theory, where the number of modes is finite and their eigenvalues 
discrete,  there are at least two seemingly natural ways to choose the upper cutoff: 
including contributions up to a given mode number (say N), or alternatively, setting a cutoff in 
the magnitude of the eigenvalues.  These two procedures lead in general to different results \cite{RvN}.
Fortunately the essence of this problem was analyzed in detail by Rebhan and 
van Nieuwenhuizen \cite{RvN}, who  showed that only a mode number cutoff leads to 
consistent results.  By including all of the $N$ modes, bound and unbound, in our numerical 
treatment of these sums, we effectively implement this prescription without difficulty. 

We are now left with the task of solving the interacting self-consistent problem 
numerically. We use a standard relaxation routine \cite{Nrecipes} to solve Eq.~(\ref{eqphi}), 
together with a set of LAPACK packages \cite{LAPACK} to solve the real symmetric 
eigenvalue problem Eq.(\ref{eqG}-\ref{chidef}). At each step  the fluctuation eigenproblem is solved
to produce a new intermediate $G_{\rm int}(x)$, which we combine with the old value to produce a new 
trial $G_{\rm new} = (1-\gamma) G_{\rm old} + \gamma G_{\rm int}$, until convergence in $\varphi$ and 
$G(x)$ is reached, up to a specified precision. 
The adjustable parameter $0 < \gamma < 1$ controls the size of the update and can be adapted to optimize 
convergence.  

Both mean-field and fluctuation update steps can be problematic and require some further discussion. 
The system retains an overall translational mode and iterative procedures where updates of $\varphi$ 
and $G$ are staggered can excite it, thus degrading convergence. To eliminate this problem we 
pin the kink's center  $x_0 :  \varphi(x_0)=0$ to the mid-point of our spatial lattice. 
The full profile for the mean field is then obtained through the symmetry $\varphi(-(x-x_0)) = -\varphi(x-x_0)$. 
To strict 1-loop order there is a well-known fluctuation zero mode associated with the eigenvector 
$g_0(x) \sim \partial_x \phi$, regardless of the specific mean-field background.  
This follows immediately by taking a spatial derivative through the (classical)  
Eq.~(\ref{eqphi})  and comparing the result to the 1-loop eigenvalue problem.
This construction ceases to apply once fluctuations are made self-consistent, as their 
potential is shifted by a term $\sim 3 \lambda G(x)$.  Neglecting back-reaction on $\varphi(x)$ 
and assuming this effect to be small, knowledge of $G(x)$ allows us to estimate the energy shift
of the former zero mode as $\omega^2_0 \rightarrow  3 \lambda \int dx ~ g^*_0(x) G(x) g_0(x)$, where $g_0(x)$ is the 
1-loop zero mode eigenvector.  This compares well with with our numerical results, even if $\lambda$ is not small.
It predicts $\omega^2_0= 0.513m^2$ {\it vs.} $\omega^2_0=0.504m^2$, with the parameters of 
Table~\ref{tab1}, shown below. Thus we see that no {\it fluctuation} zero-mode persists in 
the self-consistent problem. We must nevertheless emphasize that the dressed kink retains an 
overall translational invariance at no energy cost (as the energy functional is independent of where the 
kink is placed) but that this transformation requires a simultaneous translation of the mean-field 
profile and its self-consistent fluctuations.   

\begin{table}
\begin{tabular}{lcccc} 
\hline \hline
approximation  & Energy & $\omega^2_0$ & $\omega^2_1$ & $\omega^2_2$ \\
\hline
classical & 0.9428 &   - & - & -  \\
1-loop & 0.4716   &  0 & $\frac{3}{2} m^2$ & $2 m^2$ \\
1-loop resummed &  0.7184  &  $0.50 m^2$ & $1.71 m^2$ & $2 m^2$ \\ 
Hartree &  0.6634  &   $0.33 m^2$ & $1.29 m^2$ & $1.90 m^2$ \\ 
\hline \hline
\end{tabular}
\caption{Numerical results for the kink rest energy in the classical, strict 1-loop, 1-loop resummed 
excluding and including (Hartree) back-reaction of quantum fluctuations 
on the mean field $\varphi$. Here we have taken $N=800$, $L=20$ and $\lambda/m^2=\hbar=1$.
We also show the three lowest lying eigenvalues for each of these approximations.}
\label{tab1}
\end{table}

The advantage of studying the kink is that several analytical results are known, which  
we can use as checks on our numerical methods. The classical kink energy takes the 
simple form $E = \frac{2 \sqrt{2} }{ 3} \frac{m^3}{  \lambda}$. With our choice of parameters 
this is $E=0.9428$, which coincides with the energy of our numerical solution, 
as shown in Table~\ref{tab1}.

\begin{figure}
\vspace{0.5in}
\includegraphics[width=5in]{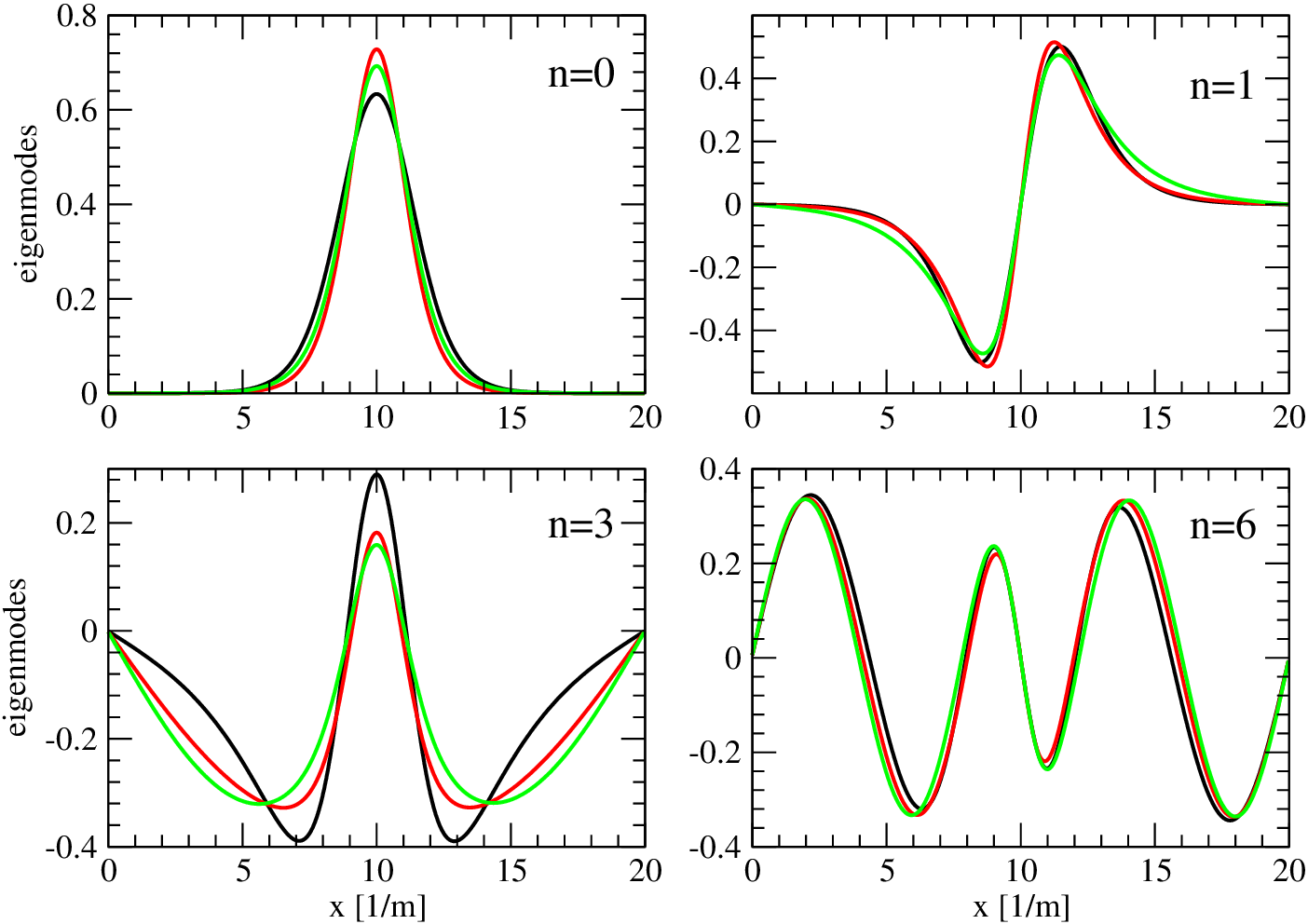}
\caption{\label{fig1}   Four of the lower eigenmodes ($n=0,1,2,6$, clockwise) in the 1-loop 
approximation (red), 1-loop resummed in $G$ (green) and full problem including backreaction 
on $\varphi$ (black).}
\end{figure}

At 1-loop order the spectrum of fluctuations about the classical kink 
is also known exactly. The spectrum consists of two discrete modes with
eigenvalues $\omega^2_0= 0$, $\omega^2_1 = \frac{3}{ 2} m^2$ lying
below a continuum with $\omega^2_k = k^2 + 2 m^2$. The corresponding
(unnormalized) eigenvectors are 
\begin{eqnarray}
&& g_0(x) =  \cosh^{-2} z ; \label{g0} \qquad  
g_1(x)  = \sinh z/\cosh^2 z,  \label{g1}\\
&& g_n(x) = e^{i k x} \left[ 3 \tanh^2 z  - 3 \sqrt{2} i 
\tanh z  - 1 - 2 \frac{k^2}{m^2} \right]. \qquad 
\end{eqnarray} 
with $z=m x/\sqrt{2}$. We verified that to this order our numerical solutions coincide with these results. 
Fig.~\ref{fig1} shows the eigenvectors obtained numerically for low lying modes. Corresponding 
eigenvalues and kink rest energies computed at different levels of approximation 
are summarized in Table ~\ref{tab1}.
The 1-loop results for the mass shift $\delta m$ computed via two different methods (\ref{E1}) and (\ref{E2}) agree well with each
other ($-0.4712$ {\it vs.} $-0.4742$) and also with the known analytical result 
at the 1-loop order, $\delta m=-0.4711$, see Refs.~\cite{Dashen,Rajaraman,BT,Weidig}. 
Beyond this point no exact analytical results are known. 

As a first step beyond 1-loop, we may consider solving the self-consistent problem
Eq.~(\ref{sfeigen}) {\em without} backreaction on $\varphi$. 
This case has been studied by Boyanovsky {\it et al.}
\cite{Boyanovsky} who devised a variational approach, based on the
knowledge of the 1-loop perturbative spectrum. They assume that a
solution for the fluctuation modes can be constructed by treating   
the inverse width of the kink potential $A$ and the asymptotic
fluctuation mass $M$ as variational parameters and using the
eigenvalue equation to determine a relationship between them. 
If we impose the constraints appropriate to our choices,
$A=m/\sqrt{2}$ and $M=\sqrt{2} m$, we can  
use the solutions of Ref.~\cite{Boyanovsky} to  predict the form of
the renormalized $G(x)$ to be  
\begin{eqnarray} 
G(x) = \frac{F(\eta)}{\cosh^2(xA)}, \ {\rm with} \quad  F(\eta) =
\frac{\eta}{2 \pi} \tan^{-1} (\eta), 
\end{eqnarray}
where $\eta=A/\sqrt{M^2 -A^2}$ and the lowest eigenvalue to be
$\omega^2_0 = M^2 - A^2$.  
With our choice of parameters this would give $\eta=0.408$, $F(\eta) =
0.025$ and  
$\omega^2_0 = 3/2 m^2$. Instead our numerical solution gives an
amplitude for $G$ of 0.226, and $\omega^2_0=0.50 m^2$, as we have
discussed above.  The reason for the discrepancy is that the 
self-consistent mode profiles, although reminiscent of their 1-loop
forms, display a different width both from the classical kink profile
and from each other. This allows them  
to considerably lower their associated eigenvalues and leads to a much larger $G$ than in 
Ref. \cite{Boyanovsky}.  In other words, the self-consistency in $G$ creates a
repulsive effect (a decrease in the depth of the attractive potential
provided by the classical kink) in (\ref{sfeigen}) and therefore leads to  
larger bound-state eigenvalues, which are the principal source of the energy shifts. As a consequence 
the quantum kink energy is now higher than at 1-loop, but still
smaller than its classical value.  Results for the first few modes are
summarized in Table~\ref{tab1}.

\begin{figure}
\vspace{0.5in}
\includegraphics[width=5in]{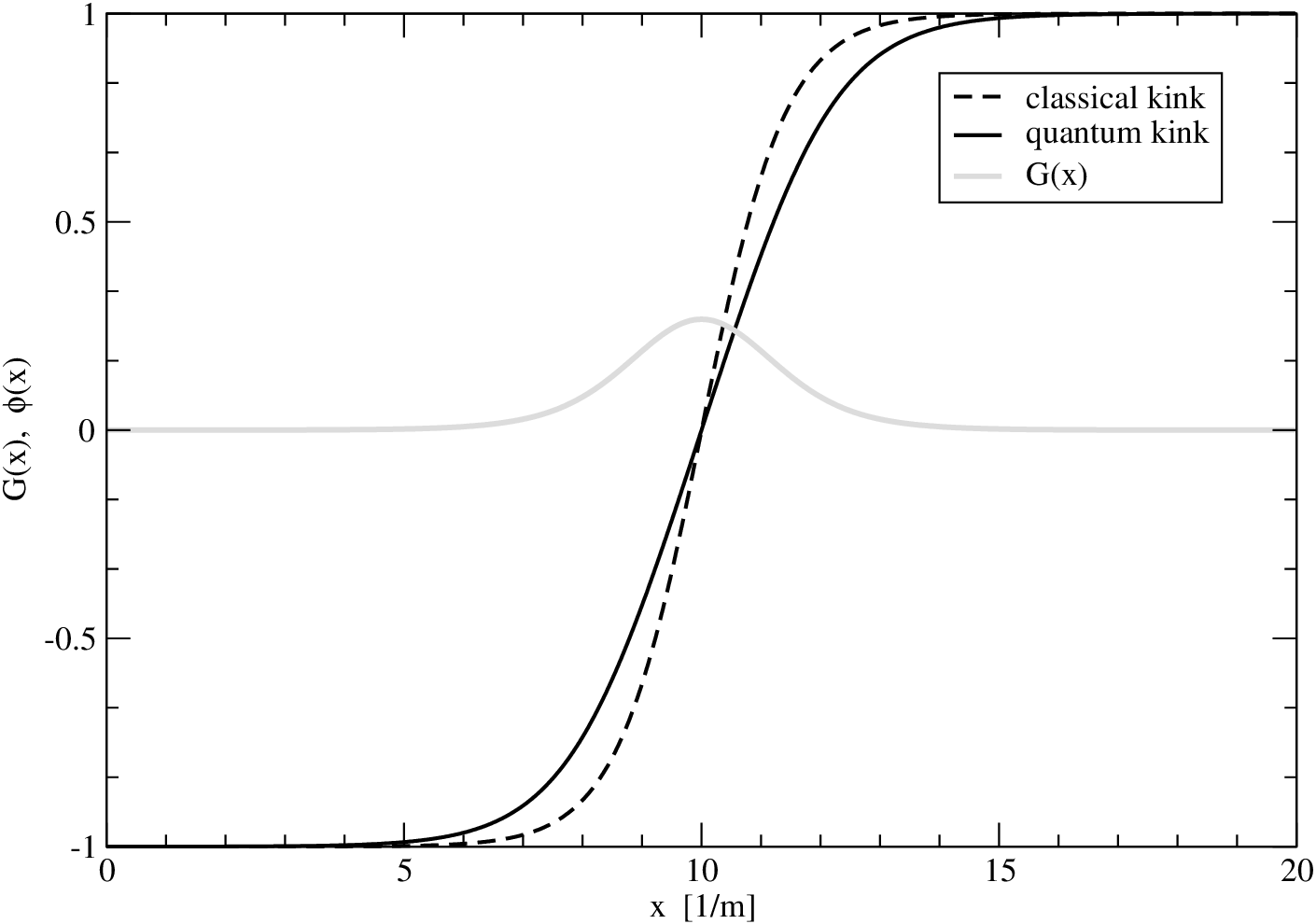}
\caption{\label{fig2} The profiles of the quantum kink, its classical counterpart and 
$G(x)$ for $\lambda=\hbar=1$}
\end{figure}

The fully self consistent solution, including backreaction on the mean-field, is 
shown in Fig.~\ref{fig2}. The most important qualitative change to the fluctuation 
spectrum is that the lower lying modes become more tightly bound relative to the 
1-loop resummed case, see Table~\ref{tab1}, and that 
a third (shallow) bound state appears. 
The kink mean-field profile is now broader than its classical counterpart.
Although the self-consistent mean-field solution is not perfectly fit by a $\tanh(x A)$, 
we can quantify this effect by producing a best value $A=m/2.07$ 
[c.f. $A=m/\sqrt{2}$ for the classical kink].  The broadening of the mean-field 
raises the classical energy but this energy cost is more than compensated 
for by the lowering of the fluctuation spectrum.    Allowing the mean-field profile 
to relax to a more favorable configuration leads to the lowering of the self-consistent 
kink energy.  Nevertheless the effects of the fluctuation self-interactions still result in 
a total energy higher than at 1-loop.  The response of the
mean-field and fluctuations taken together, self-consistently, can be
thought of as a screening effect of the defect by the
fluctuations. The mass of the self-consistent kink in this approximation is in excellent agreement
with the lattice results of Ref.~\cite{Ardekani}.

\begin{figure}
\vspace{0.5in}
\includegraphics[width=5in]{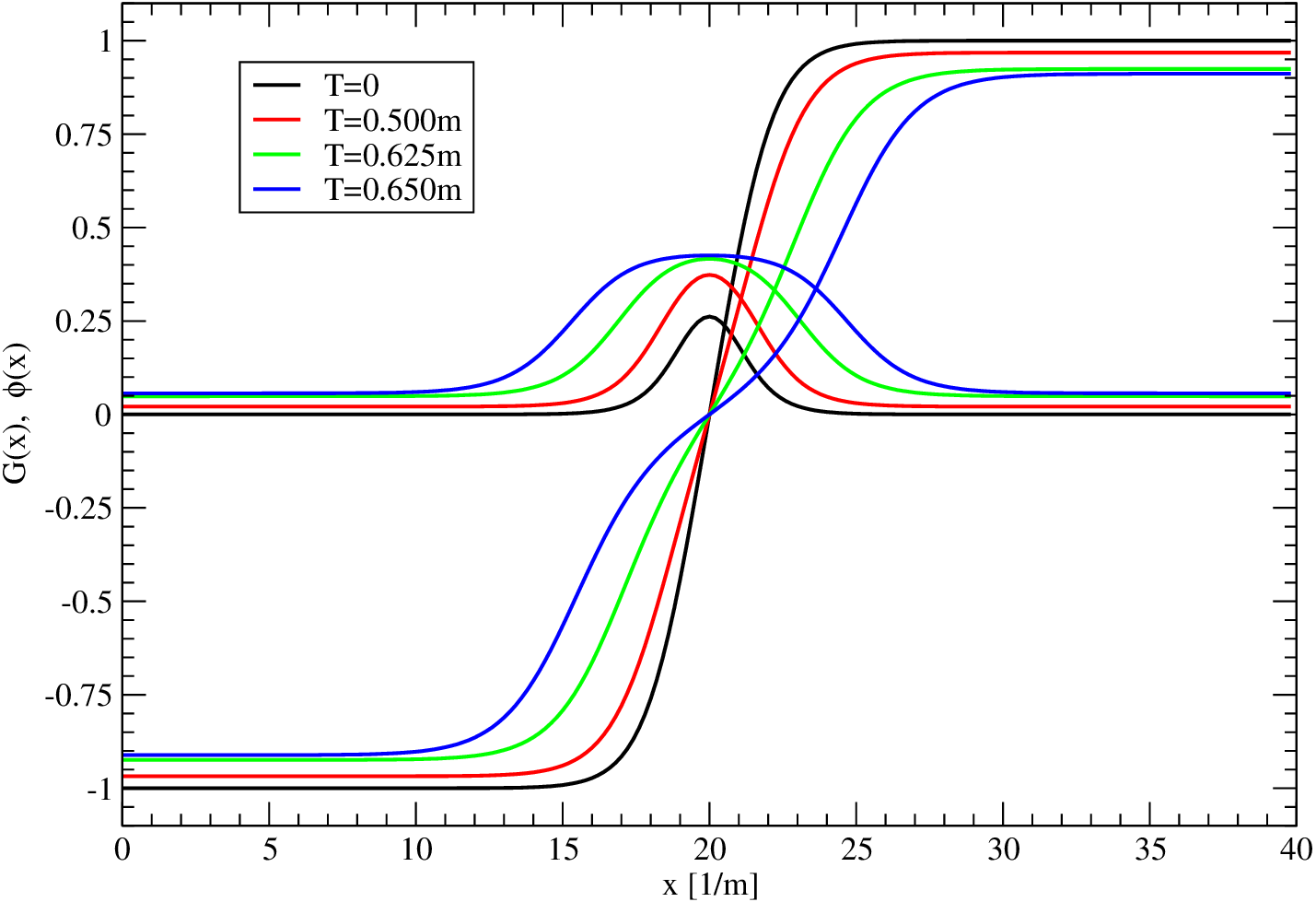}
\caption{\label{fig3}   Quantum kink profiles at different temperatures. The kink mean field profile broadens 
with temperature and eventually disappears at $T\simeq 0.65 m$. }
\end{figure}

Finally we investigate the effects of turning on temperature, see
Fig.~\ref{fig3}. Numerically this is best achieved by slowly
increasing $T$ and successively constructing solutions at higher temperature from
previous results at lower $T$. Fig.~\ref{fig3} shows that the kink broadens as the
temperature is increased.  At $T \simeq 0.65 m$ the solution starts
probing a new local minimum around $\varphi = 0$, which is an inadequacy
of the approximation (the Hartree approximation predicts a first-order
thermal phase transition for the $\phi^4$ model, which is known to
have a second order transition).  However, as the
kink energy now becomes comparable to the temperature, it is plausible
that a stable kink no longer exists; our numerical methods invariably
converge to the trivial 
$\varphi=0$ solution.  In a model with a true second order phase transition one may
expect self-consistent defects to persist and broaden all the way up to $T=T_c$, where
they may melt away (their core size diverges) and cease to be localized objects.

In summary we constructed a fully self-consistent solution for
a topological defect, including the back-reaction of quantum
fluctuations at the level of the Hartree approximation. We have shown
that at zero temperature the self-consistent dressed kink is lighter 
than its classical counterpart--the kink is still attractive and leads to a lower lying 
spectrum of fluctuations relative to the vacuum--but heavier than the well known 
1-loop result, due to the repulsion among self-consistent fluctuations.
At finite temperature the  effects of fluctuations are enhanced by a 
thermal population of low lying modes and the kink becomes ever 
broader as the temperature is increased until it vanishes.

We thank D.~Boyanovsky, F.~Cooper, R.~Jackiw, B.~Jaffe, 
K.~Rajagopal, N.~Svaiter and H.~Weigel for discussions and comments. LB benefited 
additionally from discussions with F.~Alexander, F.~Cooper, S.~Habib and E.~Mottola, 
where the idea of the present paper originated. This work was supported in part by 
the D.O.E. under research agreement $\#$DF-FC02 94ER40818 and the
DOE/BES Program in the Applied Mathematical Sciences, Contract KC-07-01-01.

\end{document}